%% file: main.tex
\documentclass[fleqn,usenatbib]{mnras}

\usepackage{newtxtext,newtxmath}

\usepackage[T1]{fontenc}
\usepackage{ae,aecompl}

\usepackage{graphicx}	
\graphicspath{{imgs/}{../imgs/}}
\usepackage{amsmath}	
\usepackage{mathtools}
\usepackage{relsize}
\usepackage{listings}
\usepackage{xcolor}
\usepackage{array}
\usepackage{bm}
\usepackage[scr=rsfso,calscaled=.96]{mathalpha}
\usepackage{subfiles}
\usepackage{wasysym}

\defcitealias{DB98}{D8}
\defcitealias{paper2}{Paper I}
\newcolumntype{L}{>{\arraybackslash}m{4.5cm}}

\newcommand{\cmmnt}[1]{}

\title[New stellar velocity substructures from Gaia DR3]{New stellar velocity substructures from Gaia DR3 proper motions}

\author[D. Mikkola, P. J. McMillan, D. Hobbs]{
Daniel Mikkola\thanks{E-mail: mikkola@astro.lu.se }$^{1}$,
Paul J. McMillan\thanks{E-mail: paul@astro.lu.se }$^{1}$,
David Hobbs$^{1}$ 
\\
$^{1}$Lund Observatory, Lund University, Department of Astronomy and Theoretical Physics, Box 43, SE-22100, Lund, Sweden}

\date{Accepted XXX. Received YYY; in original form ZZZ}

\pubyear{2022}

\begin{document}
\label{firstpage}
\pagerange{\pageref{firstpage}--\pageref{lastpage}}
\maketitle

\begin{abstract}
\input{sections/abstract}
\end{abstract}




\section{Introduction}\label{sec:intro}
\input{sections/introduction}

\section{Sample selection}\label{sec:sample}
\input{sections/method}
\section{Velocity distribution without RVs}\label{sec:theory}
\input{sections/theory}
\input{sections/results-disc}
\section{The local stellar halo}\label{sec:fv_results}
\input{sections/results-halo}
\section{Conclusions}\label{sec:conc}
\input{sections/conclusions}
\section*{Acknowledgements}
 This work has made use of data from the European Space Agency (ESA) mission {\it Gaia} (\url{https://www.cosmos.esa.int/gaia}), processed by the {\it Gaia}  Data Processing and Analysis Consortium (DPAC,  \url{https://www.cosmos.esa.int/web/gaia/dpac/consortium}). Funding for the DPAC has been provided by national institutions, in particular the institutions participating in the {\it Gaia} Multilateral Agreement. This work made use of Astropy:\footnote{http://www.astropy.org} a community-developed core Python package and an ecosystem of tools and resources for astronomy \citep{astropy1, astropy2, astropy3}.

 We thank members of Lund Observatory for helpful comments and ideas. Computations for this study were performed on equipment funded by a grant from the Royal Physiographic Society in Lund. PM is supported by research project grants from the Swedish Research Council (Vetenskapr\aa det Reg: 20170-03721 and 2021-04153). DH and PM gratefully acknowledge support from the Swedish National Space Agency (SNSA Dnr 74/14 and SNSA Dnr 64/17).
 
\section*{Data availability}
	All data analysed in this paper are publicly available from the Gaia  archive (\url{http://gea.esac.esa.int/archive/}). The 3D probability distributions used in Figures \ref{fig:disc_fv} and \ref{fig:fv_halo} are available upon request to the corresponding author.
\bibliographystyle{mnras}
\bibliography{references}


\appendix
\newpage
\onecolumn

\section{Gaia archive query}\label{app:query}
The following query has been used on the Gaia archive\footnote{\url{https://gea.esac.esa.int/archive/}} to obtain our \texttt{DISC} samples and is detailed in Section \ref{sec:sample}:\\
\begin{lstlisting}[language=SQL, deletekeywords={DEC}]
select source_id, bp_rp, phot_g_mean_mag, phot_bp_rp_excess_factor, ruwe, ra, dec,
parallax, pmra, pmdec, parallax_error, pmra_error, pmdec_error, parallax_pmra_corr,
parallax_pmdec_corr, pmra_pmdec_corr, visibility_periods_used, astrometric_chi2_al,
astrometric_n_good_obs_al, radial_velocity,
radial_velocity_error,
if_then_else(
bp_rp > -20,
	to_real(case_condition(
		phot_bp_rp_excess_factor - (1.162004 + 0.011464* bp_rp + 0.049255*power(bp_rp,2)
														      - 0.005879*power(bp_rp,3)),
		bp_rp < 0.5,
		phot_bp_rp_excess_factor - (1.154360 + 0.033772* bp_rp + 0.032277*power(bp_rp,2)),
		bp_rp >= 4.0,
		phot_bp_rp_excess_factor - (1.057572 + 0.140537*bp_rp)
	)),
	phot_bp_rp_excess_factor
) as excess_flux
from gaiadr3.gaia_source
where parallax_over_error > 10
and parallax > 5
and ruwe < 1.15
and phot_g_mean_flux_over_error > 50
and phot_rp_mean_flux_over_error > 20
and phot_bp_mean_flux_over_error > 20
and visibility_periods_used > 8
and astrometric_chi2_al/(astrometric_n_good_obs_al-5)
	  < 1.44*greatest(1,exp(-0.4*(phot_g_mean_mag-19.5)))
\end{lstlisting}

To create our \texttt{HALO} samples, we have used a similar query with a different parallax cut:\\
\begin{lstlisting}[language=SQL, deletekeywords={DEC}]
select source_id, bp_rp, phot_g_mean_mag, phot_bp_rp_excess_factor, ruwe, ra, dec,
parallax, pmra, pmdec, parallax_error, pmra_error, pmdec_error, parallax_pmra_corr,
parallax_pmdec_corr, pmra_pmdec_corr, visibility_periods_used, astrometric_chi2_al,
astrometric_n_good_obs_al, radial_velocity,
radial_velocity_error,
if_then_else(
	bp_rp > -20,
	to_real(case_condition(
		phot_bp_rp_excess_factor - (1.162004 + 0.011464* bp_rp + 0.049255*power(bp_rp,2)
														      - 0.005879*power(bp_rp,3)),
		bp_rp < 0.5,
		phot_bp_rp_excess_factor - (1.154360 + 0.033772* bp_rp + 0.032277*power(bp_rp,2)),
		bp_rp >= 4.0,
		phot_bp_rp_excess_factor - (1.057572 + 0.140537*bp_rp)
	)),
	phot_bp_rp_excess_factor
) as excess_flux
from gaiadr3.gaia_source
where parallax_over_error > 10
and parallax > power(3, -1)
and ruwe < 1.15
and phot_g_mean_flux_over_error > 50
and phot_rp_mean_flux_over_error > 20
and phot_bp_mean_flux_over_error > 20
and visibility_periods_used > 8
and astrometric_chi2_al/(astrometric_n_good_obs_al-5)
	  < 1.44*greatest(1,exp(-0.4*(phot_g_mean_mag-19.5)))
and 4.74*sqrt(power(pmra, 2) + power(pmdec, 2))/parallax > 200
\end{lstlisting}
\twocolumn

\section{Red and blue CMD sequence selection}\label{app:regions}
Table \ref{tab:regions} lists the vertices for the intersect between red and blue regions of the CMD outlined in Section \ref{sec:sample}. In addition to this, the blue section starts from the point $(G_\mathrm{BP}-G_\mathrm{RP}, M_G) = (-0.125, 11)$ mag and passes through the points (0.8, 1) mag and (-0.125, 1.95) mag after the intersecting vertices. The red section starts from (3.125, 11) mag and after the intersecting vertices ends at (3.125, 5.25) mag.
\begin{table}
	\label{tab:regions}
	\centering
	\caption{Vertices for the regions in  $M_G$ and $G_\mathrm{BP}-G_\mathrm{RP}$ that make up our red and blue halo sequences.}
	\begin{tabular}{l l}
		$G_\mathrm{BP}-G_\mathrm{RP}$ & $M_G$\\
		\hline
		$\mathrm{mag}$ & $\mathrm{mag}$\\
		\hline
		2.4 & 11\\
		2.110 & 10.087\\
		2.012 & 9.522\\
		1.789 & 8.630\\
		1.637 & 8.178\\
		1.452 & 7.565\\
		1.262 & 7.021\\
		1.142 & 6.652\\
		1.034 & 6.239\\
		0.936 & 5.826\\
		0.843 & 5.326\\
		0.756 & 4.782\\
		0.729 & 4.478\\
		0.708 & 4.000\\
		0.713 & 3.695\\
		0.756 & 3.413\\
		0.849 & 3.326\\
		0.925 & 3.260\\
		0.963 & 3.239\\
		0.990 & 2.412\\
		1.066 & 1.173
	\end{tabular}
\end{table}

\section{Halo velocity distributions in binned velocities}\label{app:binned}
In Section \ref{sec:fv_results} we discuss certain features in the concept of the full 3D velocity structure. The way we visualise the full structure is by looking at the different velocity spaces in bins of the third velocity component, here in steps of 100 km s$^{-1}$. These figures show how some of the features are related across velocity spaces.
\begin{figure*}
	\centering
	\includegraphics[width=.8\textwidth]{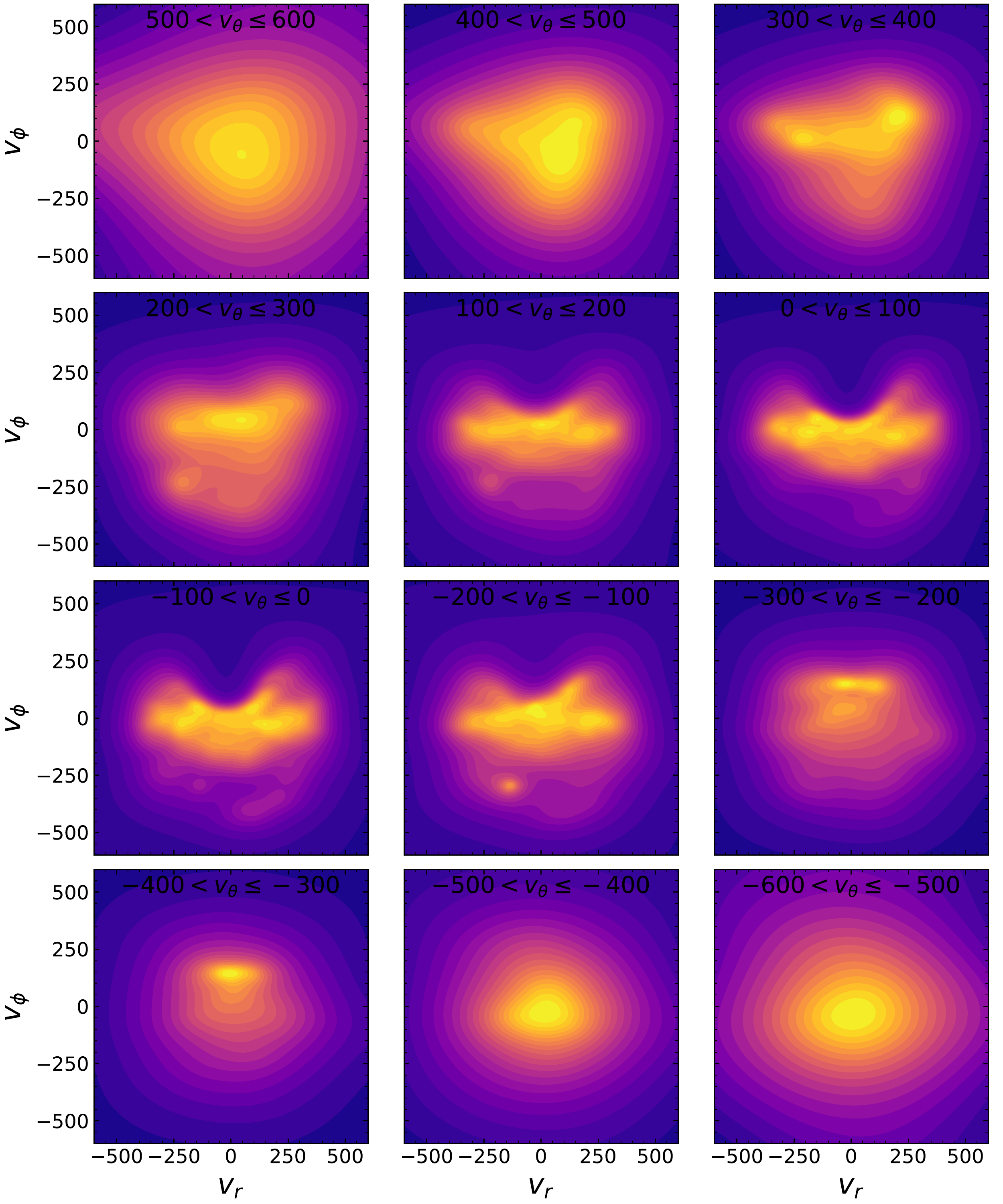}
	\caption{Velocity distributions in spherical coordinates $v_r$ and $v_\phi$, binned by their $v_\theta$.}
	\label{fig:binned_A}
\end{figure*}
\begin{figure*}
	\centering
	\includegraphics[width=.8\textwidth]{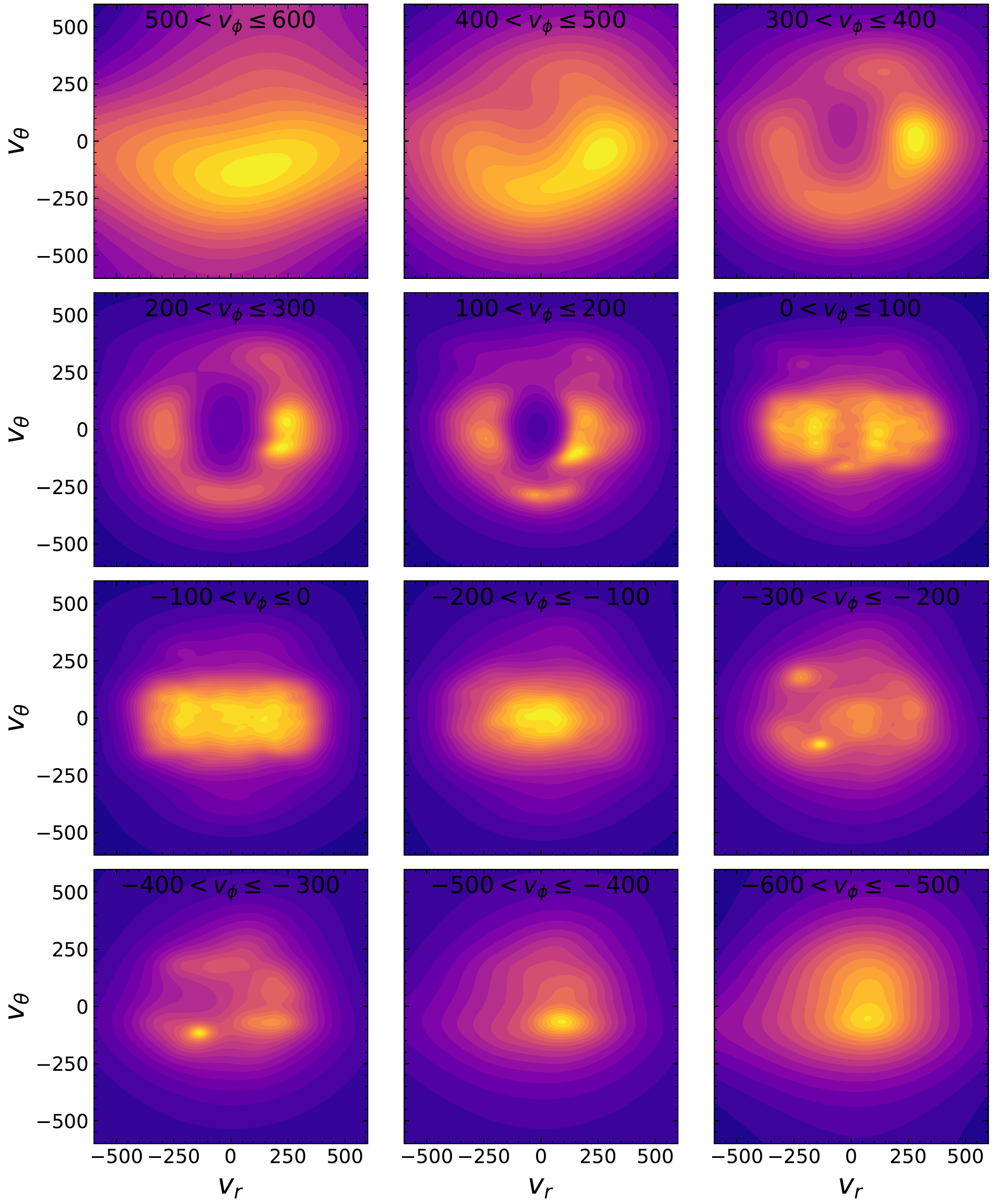}
	\caption{Velocity distributions in spherical coordinates $v_r$ and $v_\theta$, binned by their $v_\phi$.}
	\label{fig:binned_B}
\end{figure*}
\begin{figure*}
	\centering
	\includegraphics[width=.8\textwidth]{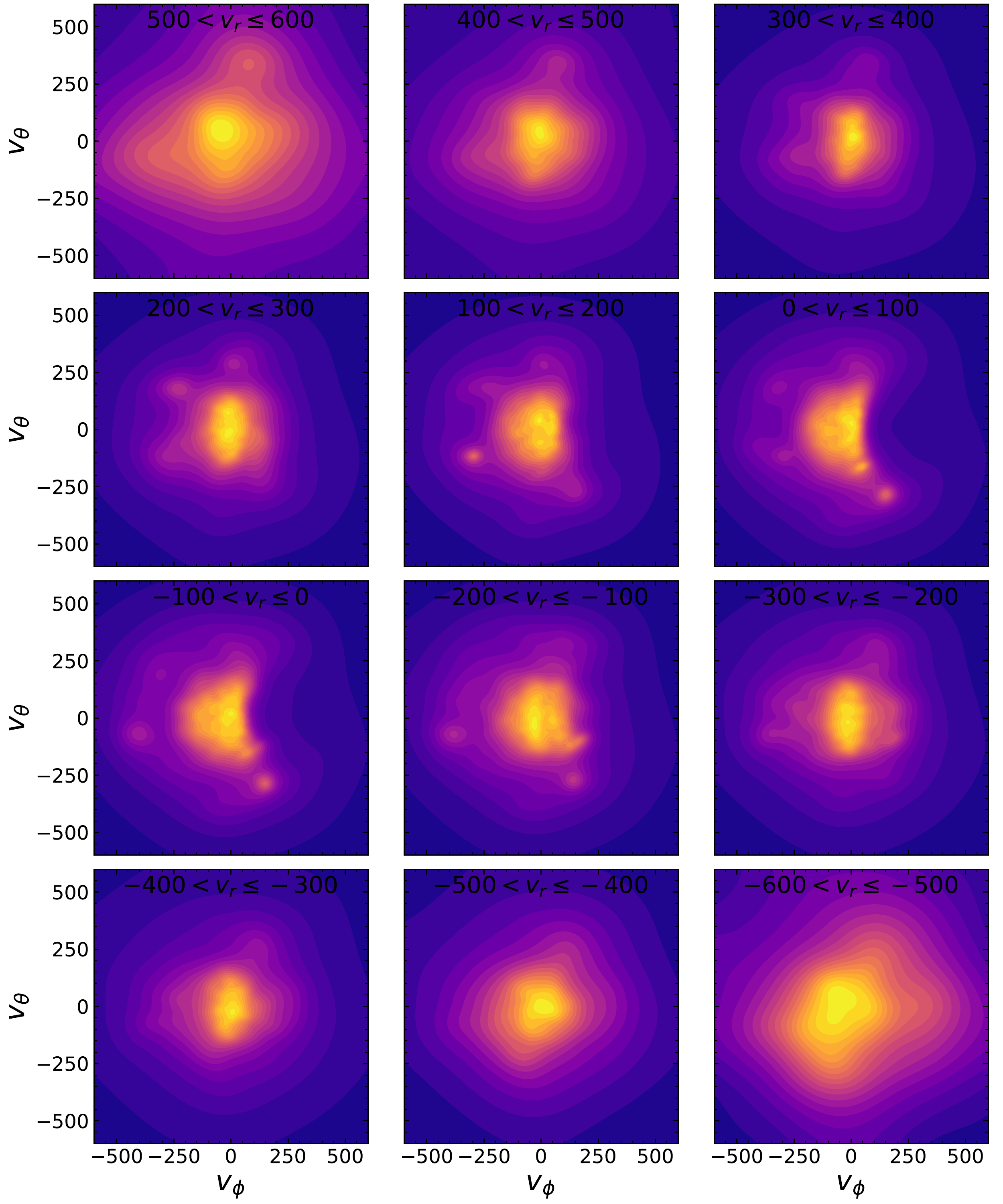}
	\caption{Velocity distributions in spherical coordinates $v_\phi$ and $v_\theta$, binned by their $v_r$.}
	\label{fig:binned_C}
\end{figure*}
\bsp	
\label{lastpage}
\end{document}

%% file: sections/abstract.tex
Local stellar motions are expected, and have been shown, to include signatures of the Galaxy's past dynamical evolution. These are typically divided into the disc, which shows the dynamical effects of spiral arms and the bar, and the stellar halo, with structures thought to be debris from past mergers. We use \textit{Gaia} Data Release 3 to select large samples of these populations without limiting them to sources with radial velocities. We apply a penalised maximum likelihood method to these samples to determine the full 3D velocity distribution in Cartesian $(U, V, W)$ or spherical $(v_r,  v_\phi, v_\theta)$ coordinates. 
We find that the disc population is dominated by four moving groups and also detect a new moving group at $(U, V) = (-10, -15)$ km s$^{-1}$ which we call \textit{MMH-0}. For the stellar halo, we isolate the accreted component with cuts in transverse velocity and the colour-magnitude diagram. In this component we find several known structures believed to be caused by past mergers, particularly one around $(v_r,  v_\phi, v_\theta) = (-150, -300, -100)$ km s$^{-1}$ appears more prominent than previously claimed. 
Furthermore we also identify two new structures near $(v_r, v_\phi, v_\theta) = (225, 25, 325)$ km s$^{-1}$ and $(0, 150, -125)$ km s$^{-1}$ which we refer to as \textit{MMH-1} and \textit{MMH-2} respectively.  
These results give new insights into local stellar motions and shows the potential of using samples that are not limited to stars with measured line-of-sight velocities, which is key to providing large samples of stars, necessary for future studies. 

%% file: sections/introduction.tex
As our Galaxy evolves, the kinematics of the stars that reside in it are imprinted by the various external and internal processes that affect it. The volume near the Sun is no different and contains footprints of possible dynamical resonances and interactions with nearby dwarf galaxies. For this reason, untangling the causes of the kinematic structure can give us vital information about the history of the Milky Way and its interactions with its nearest neighbours.

The approach to studying nearby kinematic space has bifurcated into the study of the Galactic disc (e. g. \citealt{wd98, kushniruk, antojanature, lucchini, disturbed}) and the study of the Galactic stellar halo (e. g. \citealt{koppelman19b, koppelman19a, koppelmanhelmi, lovdal, ruiz-lara_structure, dodd}). This seems only natural with the two stellar components recording different dynamical processes. The Galactic disc in the Solar neighbourhood shows evidence of dynamical resonances from the spiral arms and the bar \citep{antoja2010, trick_actions}. The stellar halo, however, records evidence of mergers between the Milky Way and its neighbours \citep[e.g.,][]{helmi2020} which in the $\Lambda$ cold dark matter ($\Lambda$CDM) model is how galaxies build up their halos. Finding the causes behind the structures that we can see in the velocity distributions of the local Galaxy will be an important step towards fully understanding its complex history.

As noted by \cite{helmi2020}, large samples with accurate kinematics are required if we are to detect each individual structure. This requirement is now starting to met by the advent of \textit{Gaia} \citep{gaia} and it's subsequent data releases: DR2 \citep{dr2}, EDR3 \citep{edr3}, and DR3 \citep{dr3}. Thanks to this, we now have over 1.4 billion sources measured with 5D phase-space coordinates: positions and proper motions. As of DR3, ${\sim}$33 million sources also have radial velocities, about 2\% of all sources, increased from 0.5\% in EDR3. Unfortunately as we start to look at more local samples, apply quality cuts, and pick out specific populations, the number of useful sources can rapidly decline. As an example consider the local ($\varpi > 1/3$ mas) stellar halo (defined as $v_\mathrm{T} > 200\ \mathrm{km s}^{-1}$) with good parallaxes ($\varpi / \sigma_\varpi > 10$). This sample will contain 503 572 sources with 5 parameters, which is reduced to only 84 784 with measured radial velocities. By working without radial velocities we are able access significantly larger datasets and important discoveries can still be reached as demonstrated by previous works using only proper motions (e.g., \citealt{DB98, wd98, antoja2017, koppelman_and_helmi, anticentre, disturbed}).

In our previous paper \cite{paper2} (hereafter referred to as \citetalias{paper2}), we implemented the penalized maximum-likelihood method of \cite{wd98} to infer the 3D velocity distribution of white dwarfs in \textit{Gaia} EDR3. We apply the same method here to an extended Solar neighbourhood sample, split into a more local stellar disc sample and a stellar halo sample. This allows us to view the velocity distribution in unprecedented velocity resolution. We particularly focus on the stellar halo which has been shown to host many structures which are likely due to the merger history of the Galaxy (see e. g. \citealt{dodd} and references therein) with \cite{naidu} even suggesting the halo could be almost entirely comprised of substructure.

The paper is organised as follows: In Section \ref{sec:sample} we describe the data selection and the quality cuts that have been made to provide a local disc-dominated sample and a stellar halo sample.  We outline some of the differences and new additions we have made to the method from \citetalias{paper2} in Section \ref{sec:theory}. Then in Section \ref{sec:disc_results} we present the velocity distribution for the stellar disc population. The velocity distribution for the stellar halo is shown in Section \ref{sec:fv_results} where we discuss each of the velocity features we see and compare to literature, as well as present new features that we identify, and discuss our findings. We finally summarise with our conclusions in Sections \ref{sec:conc}.

%% file: sections/method.tex
\begin{figure}
	\centering
	\includegraphics[width=0.50\textwidth]{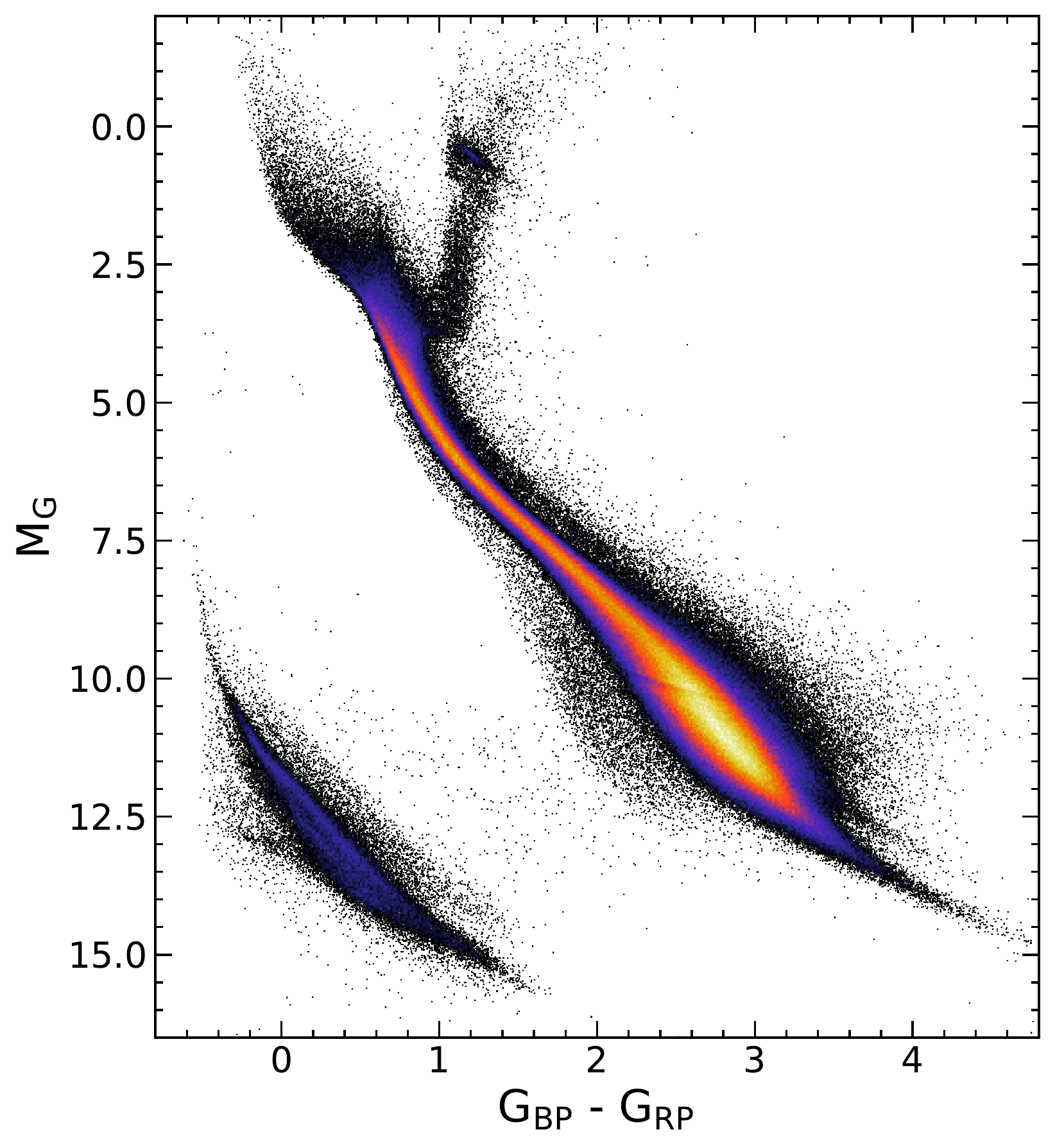}
	\caption{Colour-Magnitude diagram of our Solar neighbourhood sample. The colour shows the number density of sources. We exclude bins with N < 5 stars in them.}
	\label{fig:dr3CMD}
	\vspace{-10pt}
\end{figure}
\begin{figure}
	\centering
	\includegraphics[width=0.50\textwidth]{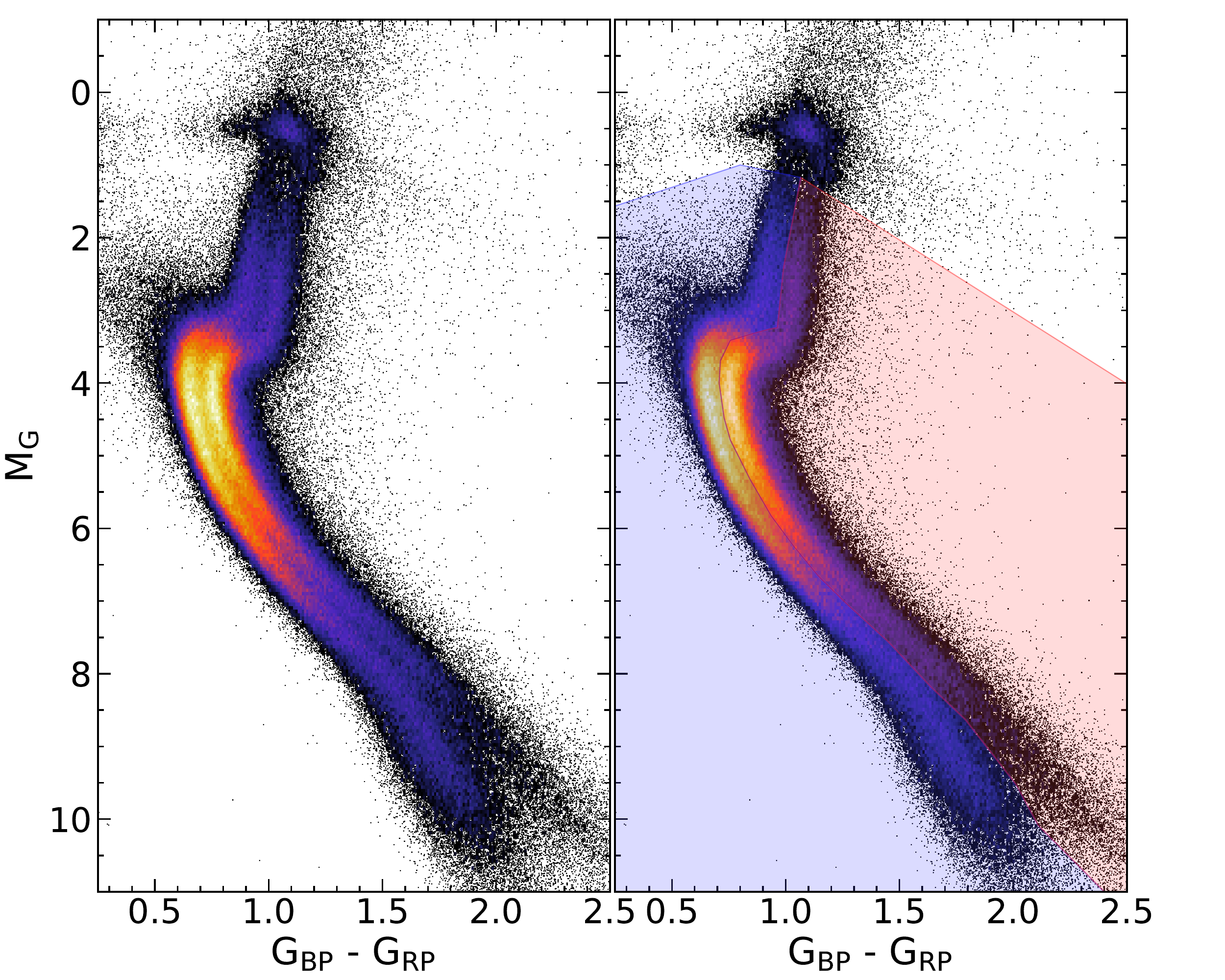}
	\caption{\textit{Left}: Colour-Magnitude diagram of our sample of stars with $V_\mathrm{T} > 200 \mathrm{km s}^{-1}$ corrected for extinction effects. The two sequences from \citet{dr2HR} are clearly visible. \textit{right}: The same plot but showing the regions used to isolate the left and right sequences in red and blue respectively.  The color shows the number density of sources and again we exclude bins with N < 5 stars in them.}
	\label{fig:haloCMD}
	\vspace{-10pt}
\end{figure}
\begin{table}
	\centering
	\caption{The names of the various samples used and the number of sources in them}
	\begin{tabular}{l | r | L}
		Name & $N$ & Description\\
		\hline
		\hline
		\texttt{DISC} & 1 171 846 & Stars in the Solar neighbourhood disc defined as $\varpi > 5$ mas  \\ \hline
		\texttt{DISC\_NORTH} & 578 368 & Same as {\tt DISC} but with $b > 0$ \\ \hline
		\texttt{DISC\_SOUTH} & 593 478 & Same as {\tt DISC} but with $b < 0$\\ \hline
		\texttt{DISC\_RVS} & 510 478 & Same as {\tt DISC} but with measured radial velocities\\ \hline
		\texttt{HALO\_BLUE} & 239 115 & Blue halo sequence, see Section \ref{sec:sample}.\\ \hline
		\texttt{HALO\_RED} & 194 507 & Red halo sequence, see Section \ref{sec:sample}\\ \hline
		{\tt HALO\_RVS} & 69 820 & Our halo sample but with measured radial velocities
	\end{tabular}
\label{tab:samples}
\end{table}
The largest sample of stars with proper motions is available from Gaia DR3 through the Gaia Archive\footnote{\url{https://gea.esac.esa.int/archive/}}. Our samples are selected to be within an extended Solar neighbourhood (Snbh, $d\lesssim$ 3 kpc) in order to meet the criteria of homogenous velocity distribution over the volume in question as outlined in Section 2.1.1 of \citetalias{paper2}. we perform a series of quality cuts, most of which are visible in our ADQL queries (which we provide in appendix \ref{app:query}). We select stars with $\varpi/\sigma_\varpi > 10$ as a strong cut on parallax uncertainties lets us approximate distance as $d = 1/\varpi$. We also filter \texttt{ruwe} $< 1.15$ to ensure good quality astrometric solutions \citep{ruwe} after inspection of the \texttt{ruwe} distribution. Additionally, \cite{dr2:astrometric} explains that for five-parameter solutions to be accepted at least six separate observations are used (called \texttt{visibility\_periods\_used} in the archive). As in \cite{dr2HR} we use a stronger filter of at least 8 which removes outliers at the fainter end. In addition, we use their criteria for \texttt{astrometric\_chi2\_al} and \texttt{astrometric\_n\_good\_obs\_al} to remove artefacts due to excess astrometric noise. Finally we use their quality filters for relative flux error on photometry:
\begin{align*}
	&\textrm{\texttt{phot\_g\_mean\_flux\_over\_error}} > 50, \\
	&\textrm{\texttt{phot\_rp\_mean\_flux\_over\_error}} > 20,\textrm{\qquad and} \\
	&\textrm{\texttt{phot\_bp\_mean\_flux\_over\_error}} > 20.
\end{align*}
\noindent Beyond this, we also calculate the flux excess in BP and RP, $C^*$, following the procedure of \cite{edr3:riello} directly in our query as in \cite{edr3}. The scatter of $C^*$ with magnitude, $\sigma_{C^*}(G)$ is fitted with a power law
\begin{equation}
    \sigma_{C^*}(G) = c_0 + c_1 G^m,
\end{equation}
where $c_0 = 0.0059898$, $c_1 = 8.817481\times 10^{-12}$, and $m=7.618399$. For each star, the value of $G$ and $C^*$ is input and we then select stars such that $C^* < 3|\sigma_{C^*}(G)|$.

Our disc sample contains stars which have $\varpi > 5$ mas ($d \lesssim$ 200 pc). This sample is further split into north and south Galactic hemisphere samples with $b > 0$ and $b < 0$ respectively. The final result of our filters on this sample can be seen in the colour-magnitude diagram (CMD) in Fig. \ref{fig:dr3CMD}. We also create a disc sample from the sources with RVs for comparison.

\begin{figure*}
	\vspace{-8pt}
	\centering
	\includegraphics[width=1\textwidth]{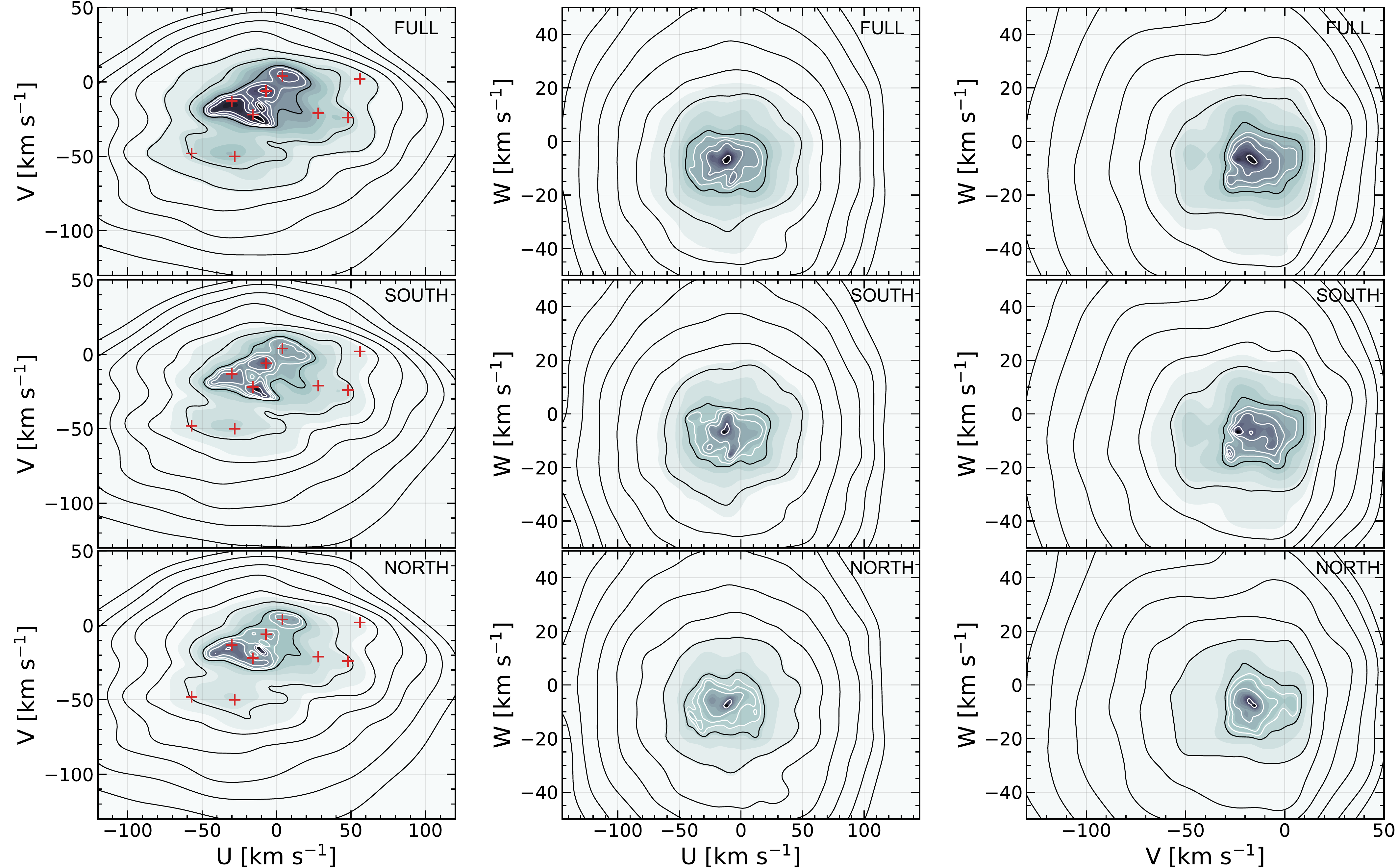}
	\vspace{-10pt}
	\caption{Galactic velocity distributions. The colormap shows the probability distribution, $f(v)$. The contour lines contain 99, 98, 96, 92, 84, 68, 35, 27, 18, 10, and 1 percent of all the stars, from outside going inward. The white contours start at 27\%. \textit{First row:} The velocity distribution of the {\tt DISC\_FULL} sample. In the first column, the location of the first 9 groups of \protect\cite{antoja2012} are shown as red crosses. \textit{Second row:} Same as the first row, but for the \texttt{DISC\_SOUTH} sample. \textit{Third row:} Velocity distribution of \texttt{DISC\_NORTH}.}
	\label{fig:disc_fv}
	\vspace{-10pt}
\end{figure*}
\begin{figure}
	\vspace{-8pt}
	\centering
	\includegraphics[width=0.45\textwidth]{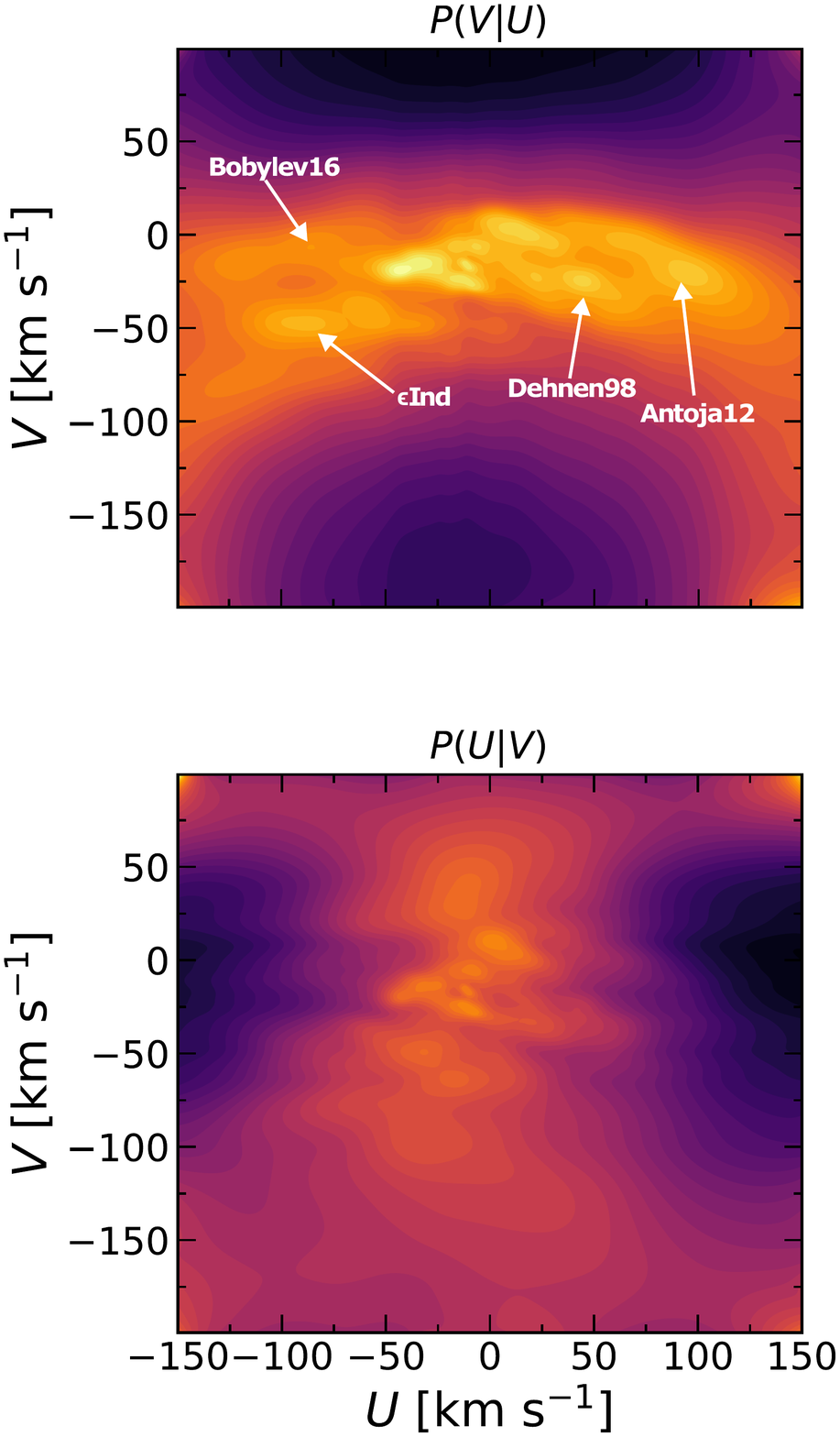}
	\vspace{-10pt}
	\caption{The conditional probability on $f(U, V)$ on either $U$ or $V$ for the top and bottom plot respectively. The density is scaled such that $P(v)^{0.25}$ to reveal low probability structure. Some features discussed in the text are highlighted.}
	\label{fig:fv_disc_conditional}
	\vspace{-10pt}
\end{figure}

To generate a sample of halo stars we select stars with a transverse velocity
\begin{equation}
    v_\mathrm{T} = \frac{4.7405}{\varpi}\sqrt{\mu_l^2 + \mu_b^2}  > 200\ \mathrm{km s}^{-1},
\end{equation}
as in \cite{dr2HR} where it was shown this reveals two dual well-defined sequences, the standard sequence of the Solar neighbourhood and a new, bluer, sequence to the left consistent with a more metal-poor population. This cut on velocity will remove some low velocity halo stars, but more importantly removes the majority of disc stars. This means our sample contains slightly fewer stars but is purer, allowing us to identify the sequences more easily. The smaller sample also makes computations less demanding. We further require $\varpi > 1/3$ mas ($d \lesssim$  3 kpc) to ensure we avoid orbits belonging to the bulge.

Our disc sample should be mostly unaffected by extinction due to its proximity and it will not have any bearing its analysis either. For the halo sample however, we need to isolate the two sequences in the CMD and therefore it is important we treat it correctly. To correct our colours and magnitudes, we use the extinction map of \cite{extinctionmap}\footnote{\url{https://stilism.obspm.fr/}} to determine the reddening for our stars . We pair this with extinction coefficients for the Gaia bands $G, G_\mathrm{BP}, G_\mathrm{RP}$ from \cite{sanders_and_das} to reduce the effects of extinction. Using the corrected values, we show the CMD in Fig. \ref{fig:haloCMD} which clearly shows the two separate sequences. Splitting the sample into a left and right sequence sample can be done reliably by eye and the selection is shown with blue and red shaded regions respectively.

We list the names of our sample as well as the number of sources contained within them in Table \ref{tab:samples}.

%% file: sections/theory.tex
As explained in Section \ref{sec:intro}, only a small minority of stars with the full astrometric solution in \textit{Gaia} DR3 also has radial velocity measurements. In the following section, we outline how we are able to estimate the full 3D velocity distribution from stars without measured radial velocities and therefore use larger samples than would otherwise be possible.

We have used the same maximum penalized-likelihood algorithm from \cite{wd98} that was used for \citetalias{paper2}. We summarise the key elements of this approach here, and a more detailed explanation can be found in our earlier paper. The method makes use of the projection of transverse velocities on the sky. The probability distribution of these transverse velocities in a specific direction, $\hat{\bm{r}}$, we write as $\rho(\bm{q|\hat{\bm{r}}})$ where $\bm{q}$ is the 2D transverse velocity. This relates to the 3D velocity distribution, $f(\bm{v})$, as:
\begin{equation}
	\rho(\bm{q|\hat{\bm{r}}}) = \int \mathrm{d}v_r f(\bm{v}) = \int \mathrm{d}v_r f(\bm{p} + v_r\hat{\bm{r}}).
\end{equation}
Here, $\bm{p}$ is the tangential motion of a star projected into 3D. Tangential motion is not sufficient to determine a true distribution and instead we must estimate it with a log-likelihood of a model for it. Numerically, we use the discrete velocity distribution
\begin{equation}
	f(\bm{v}) = e^{\phi(\bm{v})},
\end{equation}
where $\phi(\bm{v})$ is the logarithm of the probability density, which we discretize on a 3D grid of $L_U\times L_V\times L_W$ cells with widths $h_U\times h_V\times h_W$\footnote{$U$, $V$ and $W$ are the usual heliocentric velocity components in the direction towards the Galactic centre, towards Galactic rotation, and towards the north Galactic pole, respectively}. The resulting function which we use in our maximum penalized-likelihood estimation is:
\begin{multline}\label{eq:functional}
	\tilde{\mathscr{Q}}_\alpha(\bm{\phi}) = N^{-1}\sum_{k} \ln \left[\sum_{\bm{l}}e^{\phi_{\bm{l}}}K(k|\bm{l})\right] - \sum_{\bm{l}}e^{\phi_{\bm{l}}} \\ - \frac{1}{2}\alpha h_Uh_Vh_W\sum_{\bm{l}}\left(\sum_{\bm{n}} \phi_{\bm{n}}\Xi_{\bm{n}\bm{l}}\right)^2.
\end{multline}
The first term is the sum of the probability distribution function with $N$ being the sample size. For each star, $k$, the length in velocity space of the line formed by its tangential velocity and all possible radial velocities through a cell, $\bm{l}$, is $K(k|\bm{l})$. The second term is a normalising term and the third is the penalizing term where $\left(\sum_{\bm{n}} \phi_{\bm{n}}\Xi_{\bm{n}\bm{l}}\right)$ is a numerical approximation for the second derivative of $\phi(\bm{v})$ for a given cell. This term therefore penalises unsmooth solutions and is scaled by the smoothing parameter, $\alpha$.

We use a similar method for determining the optimal value of $\alpha$ as in \citetalias{paper2}. That is, we compare many estimations of an equally sized RVS sample using different $\alpha$. The comparable RVS sample to the full disc sample is only half as large as the sample without $v_r$ so we have to upscale our sample.  We create a copy of the RVS sample, \texttt{DISC\_RVS} where the Galactic positions are sampled randomly from the original. The velocities are taken from the RVS, with each proper motion and radial velocity resampled  from a multivariate Gaussian with the measured values as mean, and with uncertainties and correlation coefficients in the covariance matrix. With a randomly selected Galactic position this then transforms to Galactic velocities, $U, V, W$, and this copy is then projected back into on-sky motions with the radial velocities discarded.

For the disc sample, the upscaling only needs to be done once as the RVS is half the size of the full dataset. We then compare the estimated $\tilde{f}^\alpha_{\bm{v}}$ with the real RVS $f_{\bm{v}}$ and select the $\alpha$ that gives the smallest integrated square error (ISE)
\begin{equation}
	D(\tilde{f}^\alpha_{\bm{v}}, f_{\bm{v}}) = \int d^3 \bm{v} (\tilde{f}^\alpha_{\bm{v}} - f_{\bm{v}})^2.
\end{equation}
This gives for the full disc sample an optimal smoothing of $\alpha = 10^{-11}$ for a $\bm{n} = [304, 304, 192]$ grid over the ranges $U\in [-150, 150]\ \mathrm{km s}^{-1}$, $V\in [-200, 100]\ \mathrm{km s}^{-1}$, $W\in [-100, 100]\ \mathrm{km s}^{-1}$, corresponding to a resolution of ${\sim}1\ \mathrm{km s}^{-1}$. The same setup is used for the north and south Galactic hemisphere samples to ensure the differences are not by construction. This will however result in a slightly under-smoothed distribution for the smaller samples.

For the stellar halo, the \texttt{HALO\_RVS} sample is about 3 times smaller than the two samples \texttt{HALO\_BLUE} and \texttt{HALO\_RED} and we upscale it to three times its original size. We also no longer use heliocentric Cartesian velocity coordinates, $U, V, W$ but instead use Galactocentric spherical velocities, $v_r, v_\phi, v_\theta$ defined such that for a star in the galactic plane at the position of the Sun, $v_\phi$ and $v_\theta$ are in the same direction as V and W, respectively, to make comparisons easier. To make the transformation, we have assumed the Sun's position to be $(R, z) = (8122, 20.8)$ pc \citep{gravity2018, bennett2019} and its velocity $(U, V, W) = (12.9, 245.6, 7.78)$ km s$^{-1}$ \citep{drimmel2018} with respect to the Galactic centre. On a grid of $\bm{n} = [240, 240, 240]$ with $v_r, v_\phi$, and $v_\theta$ all in the range $[-600, 600]\ \mathrm{km s}^{-1}$ corresponding to a resolution of $5\ \mathrm{km s}^{-1}$, the value of $\alpha$ that minimises the ISE is $4.64\times 10^{-13}$ and is used for both halo samples.

One difference between our approach when handling the disc sample and when handling the halo sample is how the velocity dispersion, $\bm{\sigma}$, and average velocity, $\langle\bm{v}\rangle$, are determined. For the former sample we can determine $\bm{\sigma}$ and $\langle\bm{v}\rangle$ of the sample directly, following the procedure of \cite{DB98} (as in \citetalias{paper2}), but for the latter case we use spherical coordinates and it becomes unnecessarily complex. Instead, we determine $\langle\bm{v}\rangle$ and $\bm{\sigma}$ for the subset of the sample that has measured radial velocities. However, since $\bm{\sigma}$ and $\langle\bm{v}\rangle$ are only used for the initial guess of $\phi(\bm{v})$ and as scaling factors for determining $\alpha$, we are free to find $\sigma$ however we wish without significant consequences for the analysis.

%% file: sections/results-disc.tex
\section{The \textbf{stellar disc}}\label{sec:disc_results}
\begin{table}
	\centering
	\caption{Average velocities of moving groups mentioned in the text, based on reported values in \protect\cite{kushniruk}.}
	\begin{tabular}{l l}
        \hline
		$(\langle U\rangle, \langle V\rangle)$ & Moving group \\
        $[\mathrm{km\ s}^{-1}]$ &  \\
		\hline
		$(8, 3)$ & Sirius  \\ 
		$(-7, -8)$ & Coma Berenices \\ 
		$(-37, -17)$ & Hyades \\ 
		$(-15, -22)$ & Pleiades \\ 
		$(-35, -48)$ & Hercules \\ 
		$(46, -24)$ & Dehnen98 \\ 
        $(31, -24)$ & wolf630 \\ 
		$(-86, -46)$ & $\epsilon$Ind \\ 
		$(-95, -8)$ & bobylev16 \\
        \hline    
	\end{tabular}
\label{tab:movinggroups}
\end{table}
The velocity distribution estimated for the Solar neighbourhood sample is seen in Fig. \ref{fig:disc_fv}. The first row shows distribution for the complete sample and the second and third rows show the south and north Galactic hemisphere samples respectively. We find that the distribution is mostly dominated by the common features: \textit{Sirius}, \textit{Coma Berenices}, \textit{Hyades}, \textit{Pleiades}, \textit{Hercules}. We can also identify \textit{Dehnen98} and \textit{Wolf 630} to an extent. The first four of the major groups occupy a region that contains roughly 35\% of all the stars in the sample, shown by the white contour lines. We can see the incomplete vertical face-mixing of \textit{Coma Berenices} \citep{Quillen18, monari, bernet} as it is a much stronger feature in the southern Galactic hemisphere. A curious feature that appears more strongly in the northern hemisphere is the rather strong overdensity between \textit{Pleiades} and the expected position of \textit{Coma Berenices} at roughly $(U, V) = -10, -15\ \mathrm{km s}^{-1}$. This feature is clearly separate from \textit{Pleiades} and we find no match for it in the list of moving groups in \cite{antoja2017}, \cite{kushniruk}, or \cite{lucchini}. The close proximity to \textit{Pleiades} suggests that this feature is now visible thanks to the improved velocity resolution. When looking for this feature in the 6D sample, we find that pleiades appears to stretch toward this region but there is no separate feature. For this reason we tentatively names this a new velocity feature which we call \textit{MMH-0}. Overall we find limited detailed substructure in the direct velocity distribution which shows the dominance of the major moving groups in this space.

\subsection{Conditional \texorpdfstring{$f(\bm{v})$}{v} of the \textbf{stellar disc}}
To unravel low-level structure that the representations of the velocity distribution of Fig. \ref{fig:disc_fv} may have missed, in Fig. \ref{fig:fv_disc_conditional} we renormalize the plots so that, rather than showing the full 2D probability density of $U$ and $V$, we show the \emph{conditional} probabilities of $V$ or $U$ for each $U$ or $V$, respectively. That is, the colour represents the probability of the star having a specific $V$ given that it has certain $U$ velocity (or vice versa). 

In addition to the structure we have seen above, we can see in $P(V|U)$ around $(U, V) = -100, -50\ \mathrm{km\ s}^{-1}$ a structure that matches well with estimates of \textit{$\epsilon$Ind} (e.g., \citealt{antoja2012, bobylev:16, kushniruk}). Above it, closer to $V = -10\ \mathrm{km\ s}^{-1}$ is another feature that matches to a group identified by \cite{bobylev:16} and \cite{kushniruk}. There are also features with positive $U$ sitting at $V \approx -30$  km s$^{-1}$ with $U = 50$ km s$^{-1}$ and $U = 100$ km s$^{-1}$, the first of which is the \textit{Dehnen98} group from \cite{antoja2012}, which itself is from \citealt{wd98}. The second group is likely \textit{Antoja12} (see \citealt{kushniruk} and references therein). This demonstrates the strength of plotting conditional probabilities for the inferred velocity distributions to gain insight into low-level structures.

%% file: sections/results-halo.tex
\begin{figure*}
	\vspace{-8pt}
	\centering
	\includegraphics[width=1\textwidth]{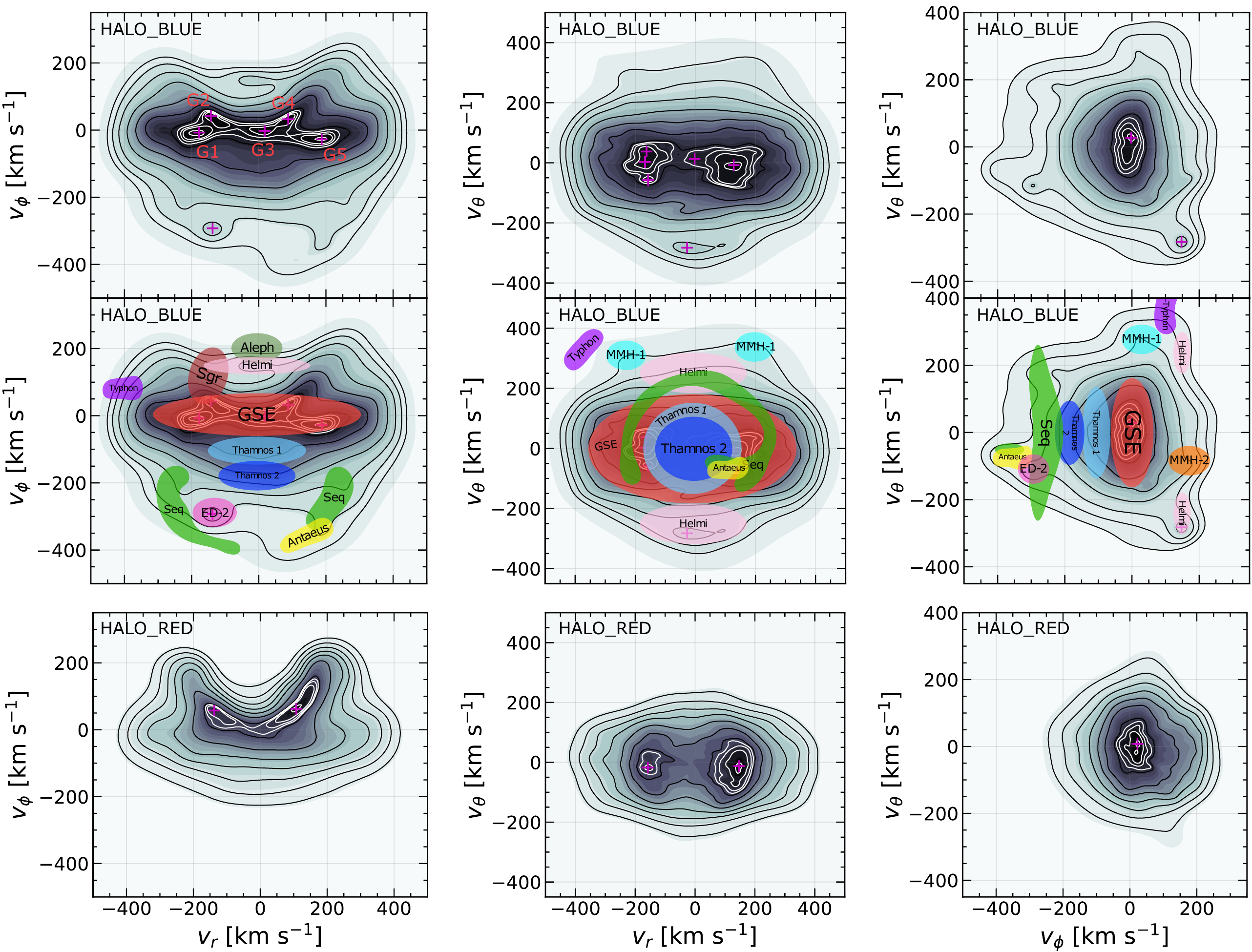}
	\vspace{-10pt}
	\caption{Velocity distributions in spherical coordinates. The colormap shows the square root of the probability distribution, $\sqrt{f(v)}$, to enhance fainter structure. The contour levels are the same as in Fig. \ref{fig:disc_fv}. \textit{First row:} The velocity distribution of the \texttt{HALO\_BLUE} sample. Five distinct features thought belonging to the region occupied by GSE are labelled. \textit{Second row:} Same as the first row, but overlaid with rough expected positions of reported substructures from literature in similar style to \protect\cite{naidu} and \protect\cite{atari} but in velocity space. \textit{Third row:} Velocity distribution of \texttt{HALO\_RED}, with very little substructure. Because we are interested in the substructure found in \texttt{HALO\_BLUE} we focus exclusively on this sample in subsequent figures.}
	\label{fig:fv_halo}
	\vspace{-10pt}
\end{figure*}
\begin{figure*}
	\vspace{-8pt}
	\centering
	\includegraphics[width=0.98\textwidth]{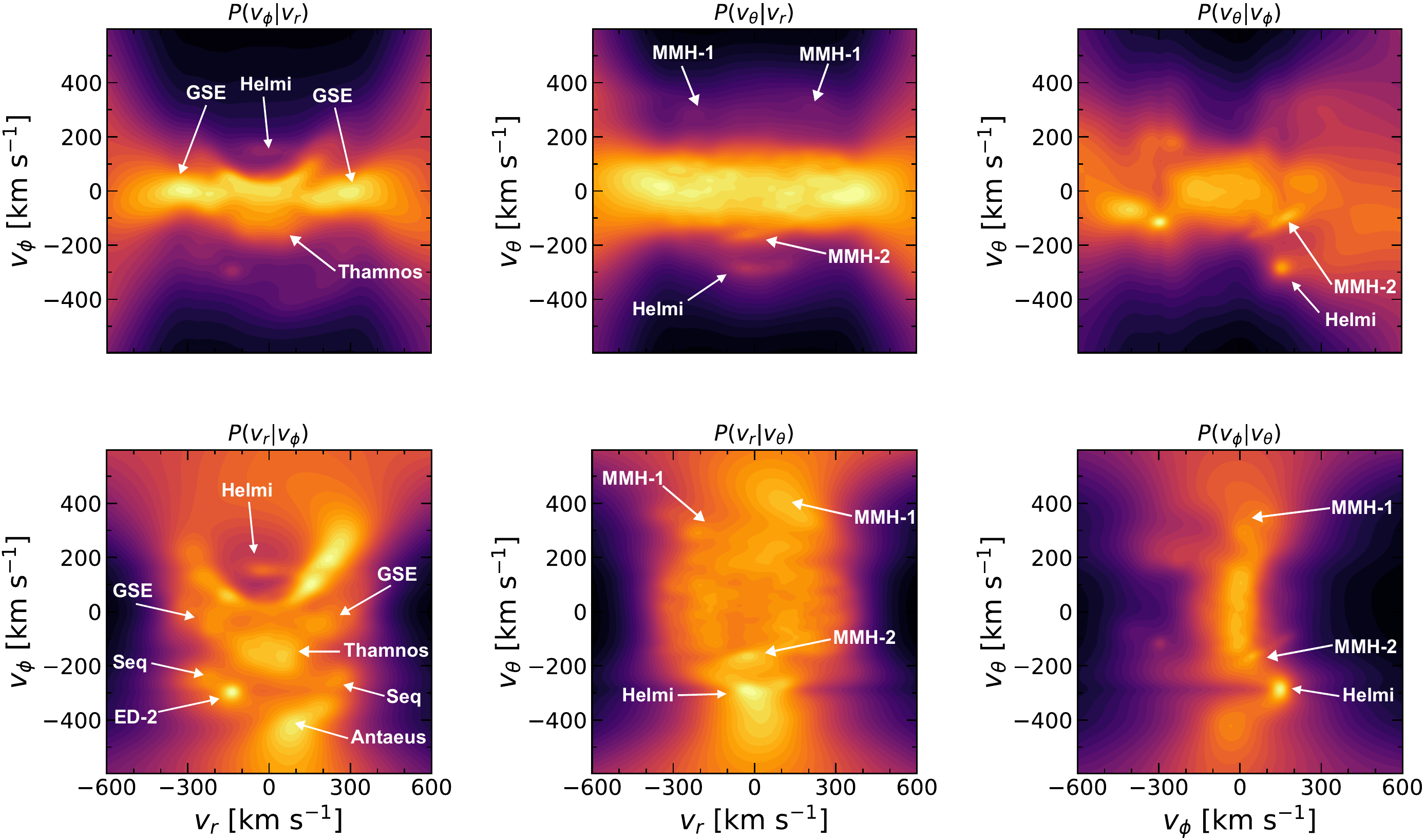}
	\caption{The conditional probability of the three different spaces in each column and velocity space as Fig. \ref{fig:fv_halo} for the \texttt{HALO\_BLUE} sample. Top row is conditional probability on the x-axis coordinate and bottom row on the y-axis coordinate. We show specifically $P(v)^{0.25}$ to reveal low probability structure even further.}
	\label{fig:fv_halo_conditional}
	\vspace{-10pt}
\end{figure*}
We present the velocity distribution of the stellar halo in the planes  $(v_r, v_\phi)$, $(v_r,  v_\theta)$, and $(v_\phi, v_\theta)$ in Fig. \ref{fig:fv_halo} estimated through our penalized maximum-likelihood method. The figure very clearly shows the separation into a blue accreted population and a red in-situ halo or hot thick disc. The blue accreted population occupies phase space almost symmetrically around $v_\phi = 0$ and does not continue smoothly from the disc population, with multiple substructures. The red, in-situ halo or hot thick disc on the other hand has very little substructure and is mostly a continuation of the disc phase space distribution. 

Since our main interest is in the substructure of the accreted halo, throughout the rest of this paper we only look at the \texttt{HALO\_BLUE} sample.

For this sample we can see that there are a plethora of features. We overlay the distribution with known features from literature, revealing which of these we do or do not see from this perspective of the population. The extent and distribution of these shapes has been adapted from visual inspection of the features as seen in \cite{koppelman19b, naidu, dodd, antaeus, typhon} and are not exact but simply to be used as a guiding tool. We will now go over and discuss the features we can see in our distributions.

\subsection{Gaia-Sausage-Enceladus}
The strongest feature across all velocity space is the \textit{GSE} and this feature is not continuous but rather appears to be a composition of multiple different features which we have labelled in the top-left of Fig. \ref{fig:fv_halo} as G1-5 in  $(v_r, v_\phi)$. Since our cut of $v_\mathrm{T} > 200\ \mathrm{km\ s}^{-1}$ is rather generous, we expect there to be some contamination between the samples. Thus, G2 and G4 are likely contaminants from the two features seen in the \texttt{HALO\_RED} sample in the third row of Fig. \ref{fig:fv_halo}. G1 and G5 appear in the space of what is typically associated with the GSE (e. g. \citealt{koppelman19b,Feuillet2021,dodd}) while the central group, G3, appears likely to be the \textit{L-RL3} group in \cite{dodd} who identifies it with Cluster 3 in \cite{lovdal}. The GSE appears slightly asymmetric on either side of $v_\phi=0$ here because of our cut on tangential velocity removing some of its lower $v_\phi$ members and including some contaminants from the disc.

\subsection{Sequoia \& Antaeus}\label{subsec:seq}
Some noticeable features are the `horns' sticking out at the bottom in $(v_r, v_\phi)$ at $v_\phi = -250\ \mathrm{km\ s}^{-1}$ and around $v_r = \pm200\ \mathrm{km\ s}^{-1}$ and to the left in $(v_\phi, v_\theta)$ around $v_\theta = \pm150\ \mathrm{km\ s}^{-1}$. 
These features are in the location associated with \textit{Sequoia} \citep{sequoia}. In \cite{naidu} this region is mixed with groups \textit{Arjuna} and \textit{I'itoi} which are distinguished from one another by their metallicities. In \citet{ruiz-lara_structure}, three separate clusters are also identified as belonging to Sequoia but share similar metallicity. Instead the kinematics distinguish these three structures and they attribute one of the clusters to the metal-poor end of the GSE based on its kinematics. Since we do not have metallicity measurements here we will refer to the dynamical space occupied by all of these features simply as \textit{Sequoia}.

In the same space we can also see the \textit{Antaeus} group from \cite{antaeus} which shares many of the same attributes as \textit{Sequoia}. It is not clear to what extent these features are separate but \cite{antaeus} claims that the low $J_z$ and position in the disc plane of Antaeus are unique.

\subsection{Helmi streams}
Two of the most prominent substructures that appears in all the distributions are the \textit{Helmi streams} \citep{helmi99}. They are also among the first identified substructures. An updated view of the streams lets us narrow them down in velocity space (\citealt{koppelman19b, koppelman19a, koppelmanhelmi}). This feature is particularly strong and is bimodal in $v_\theta$, with the lower $v_\theta$ group being far more represented as expected (e.g., \citealt{koppelman19a}). At slightly larger $v_\phi$ we would expect to find \textit{Aleph}, reported first in \cite{naidu}, but in our sample it appears to be absent. Similarly in \cite{lovdal} the absence of \textit{Aleph} was noted and argued to be caused by the velocity cut removing almost all of its stars from the sample. The same is likely the cause behind its absence in our sample.

\subsection{Thamnos}
The structure \textit{Thamnos} was identified by \cite{koppelman19a} using the \textit{Gaia} DR2 RVS sample supplemented with line-of-sight velocities and abundances from RAVE \citep{ravedr5}, APOGEE \citep{apogeedr14}, and LAMOST \citep{lamost}. Their sample was limited to 3 kpc like halo samples. In our distributions there is no distinct separate feature corresponding to \textit{Thamnos}, but in both $(v_r, v_\phi)$ and $(v_\phi, v_\theta)$ the distribution's densest parts extends to slightly lower $v_\phi$ than for the \textit{GSE} structure, which may be due to the presence of Thamnos. It is suggested by \cite{naidu} that \textit{Thamnos} may be more discernible at larger distances where the \textit{GSE} and disc-like stars contribute less to the distribution.

\subsection{Other structures}
In addition to the structures mentioned above we find several others in these projections. One of these can be associated with a known velocity structure, \textit{ED-2} from \cite{dodd} and is marked with a cerise-shaded region. This feature is close to Sequoia at $(v_r, v_\phi) = (-150, -300)\ \textrm{km s}^{-1}$ and presents in $(v_\phi, v_\theta) = (-300, -100)\ \textrm{km s}^{-1}$. To verify if both velocity representations are \textit{ED-2}, we investigate the full 3D velocity structure to find that the features overlap and are one and the same. In \cite{dodd} there is limited metallicity information for the cluster and it has average metallicity $\langle\mathrm{[Fe/H]}\rangle = -2.05$, which would match the metal poor part of \textit{Sequoia}. In Section \ref{subsec:seq} we discussed the different parts of \textit{Sequoia} which we are not able to distinguish. It is possible that this feature is yet another part of the same velocity feature.

The cluster is assigned 33 members in \cite{dodd} out of a sample of 72 274 stars (or ${\sim}$0.05\%). We can look at the fraction of the probability density that occupies the region. We define the region around the group with $v_r \in [-175, -100]$ km s$^{-1}$, $v_\phi \in [-325, -275]$ km s$^{-1}$, and $v_\theta \in [-100, -150]$ km s$^{-1}$ and find that ${\sim}$0.074\% of the sample lies there, corresponding to ${\sim}$180 stars in the sample. This suggests that the very dense feature is slightly more prominent than previous believed.

Above the high-$v_\theta$ part of the Helmi stream, at $v_\theta \approx 300\ \textrm{km s}^{-1}$, we find a new feature split across two different values of $v_r$, one around $(v_r, v_\theta) = (-250, 300) \ \textrm{km s}^{-1}$ and the other at $(v_r, v_\theta) = (200, 350)\ \textrm{km s}^{-1}$. At such a large $v_\theta$ we also find a new structure in $(v_\phi, v_\theta)$ at $(25, 300)\ \textrm{km s}^{-1}$. We refer to these groups as a single feature which we call \textit{MMH-1}. In the full 3D probability distribution the positions of \textit{MMH-1} overlap and we consider them the same feature.

Lastly, at $(v_\phi, v_\theta) = (150, -100)\ \textrm{km s}^{-1}$, another feature can be seen. At such values of $v_\phi$ and $v_\theta$ it is difficult to discern any stronger feature in the spaces of $(v_r, v_\phi)$ and $(v_r v_\theta)$ as the region is crowded particularly in $(v_r, v_\phi)$ where it would lie close to the cutoff caused by our tangential velocity limit. We refer to this feature as \textit{MMH-2}.

\subsection{Conditional \texorpdfstring{$f(\bm{v})$}{v} of the local halo}
In much the same way as we did in Fig. \ref{fig:fv_halo}, we again use conditional probabilities to illustrate our halo velocity maps with respect to one of the two velocity dimensions to investigate faint structure that otherwise may not be visible. We show these conditional probability maps in Fig. \ref{fig:fv_halo_conditional} which reveal more of the surrounding velocity structure with certain features become strikingly visible. For example the two-pronged structure around \textit{Sequioa}, \textit{Antaeus}, and \textit{ED-2} is much more readily apparent in $P(v_r | v_\phi)$ and $P(v_\theta | v_\phi)$ than before. The separation of the \textit{GSE} from the hot thick disc is apparent in both $P(v_r | v_\phi)$ and $P(v_\phi | v_r)$, and it is located primarily around two features around $v_\phi \pm [150, 300]$ km s$^{-1}$. The region occupied by \textit{Thamnos} is now also readily apparent where it was not before.

Our two novel groups, \textit{MMH-1} and \textit{MMH-2}, appear in these figures as well. The first, \textit{MMH-1}, which has a large $v_\theta$ of around 300 km s$^{-1}$ appears in the conditional probabilities $P(v_r|v_\theta)$ and $P(v_\phi | v_\theta)$ corresponding to the bottom row, middle and last columns. It also stretches up towards much larger extents in $v_\theta$, but it is unclear if this structure is physical given how far out in the edges of the distribution it lies. Similarly there is a symmetric feature at large negative $v_\theta$ near -400 km s$^{-1}$ which, if real, could relate to \textit{MMH-1}. The group \textit{MMH-2} appears strongly in $(v_\phi, v_\theta)$ for both $P(v_\phi | v_\theta)$ and $P(v_\theta | v_\phi)$.

A feature we have not discussed previously that is present in $P(v_\theta | v_r)$ and $P(v_r | v_\theta)$ is the sloped feature around $(v_r, v_\theta) = (0, -150)$ km s$^{-1}$ which we associate with \textit{MMH-2}. This feature is also visible in Fig. \ref{fig:fv_halo} and does not appear in the other papers we have reviewed and is very difficult to find in the other spaces, suggesting it lies at rather low $v_\phi$ as this would place it close to the densest parts of the distribution, thus obscuring it from detection. We have confirmed this by limiting the $(v_r, v_\theta)$-space to separate bins of $v_\phi$, which reveals that the feature only appears between $v_\phi \in [0, 100]$ km s$^{-1}$ (this can be seen in our binned velocity distributions in Appendix \ref{app:binned}). This means the feature is most likely the representation of \textit{MMH-2} in this space.

\subsection{Action space distribution}\label{sec:actions}

\begin{figure}
	\vspace{-8pt}
	\centering
	\includegraphics[width=0.45\textwidth]{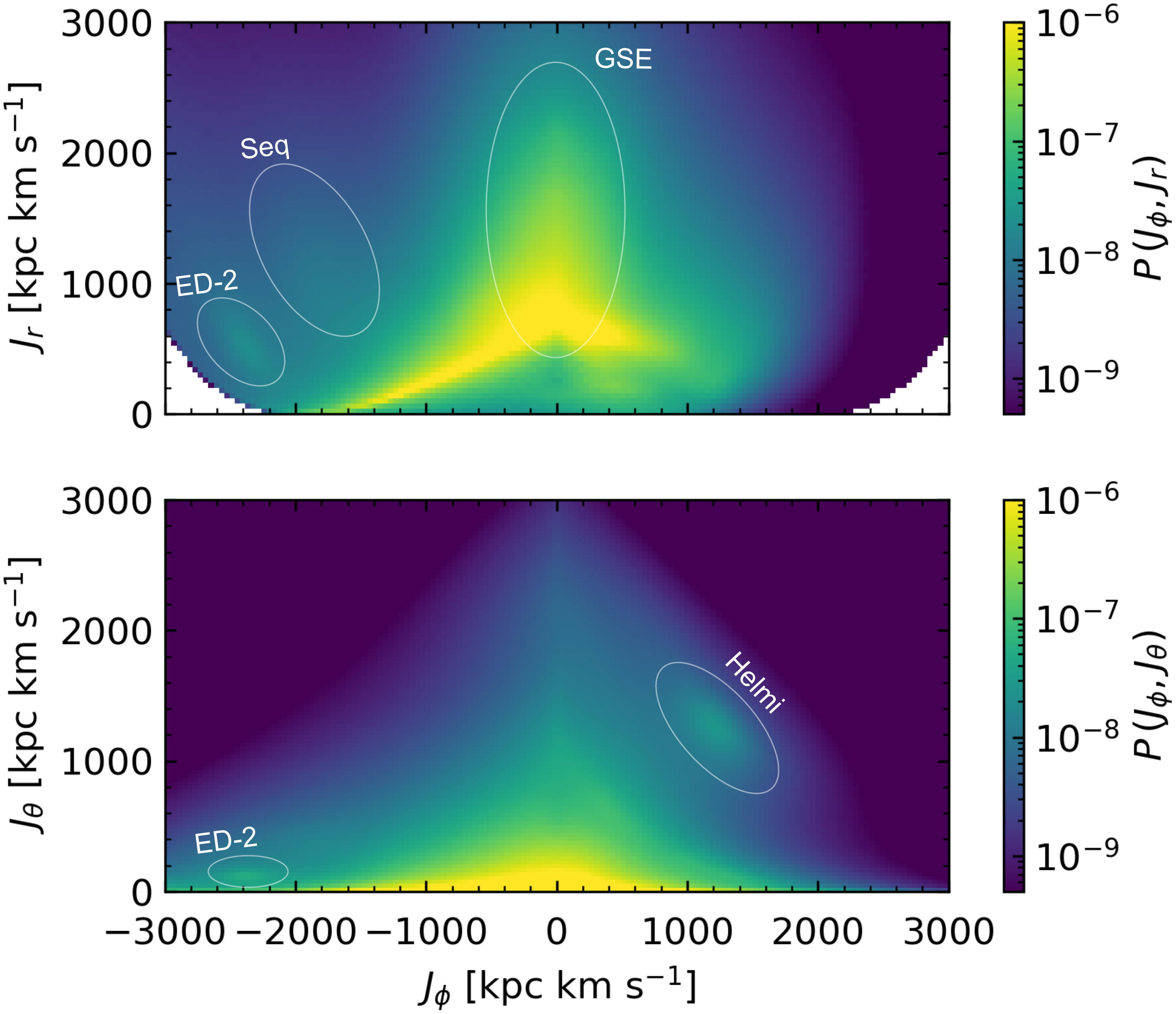}
	\vspace{-10pt}
	\caption{Action distribution associated with our \texttt{HALO\_BLUE} sample. The probability density is in units of $\mathrm{kpc}^{-2}\;\mathrm{km}^{-2}\,\mathrm{s}^2$. While there are strong selection effects that shape these distributions, we are still able to see some of the substructure in this projection of the data, most strikingly the Helmi stream in the lower panel near $(J_\phi, J_\theta)=(1300,1200)\;\mathrm{kpc}\,\mathrm{km}\,\mathrm{s}^{-1}$.}
	\label{fig:actions_halo}
	\vspace{-10pt}
\end{figure}

A very common method of searching for accreted substructure in the Milky Way halo is to work in terms of integrals of motion \citep[e.g.,][]{helmi99}. In particular, the orbital actions have become commonly used since a convenient approximation within galaxy potentials became available \citep{Binney2012}. When stars in a sample are distributed over a large volume in the Milky Way it is essential to use techniques like these to identified substructure accreted long ago because, while we can expect stars accreted together to have very similar integrals of motion, they will have very different velocities if they are in very different parts of the Galaxy.

We have very deliberately limited our sample to a relatively local volume within the stellar halo so that stars on similar orbits have similar velocities. Nonetheless, to make consistent comparison with other studies possible, it is still valuable to determine the distribution of the stars in terms of their orbital actions. We use the \textsc{agama} software package \citep{Vasiliev2019} to determine the actions in the Milky Way gravitational potential from \cite{McMillan2017} rescaled such that the Sun's position and velocity are consistent with the values given in Section \ref{sec:theory}. Fig. \ref{fig:actions_halo} shows this action distribution in terms of $J_r$ or $J_\theta$ against $J_\phi$,\footnote{$J_\phi$ is the conserved component of angular momentum, while $J_r$ characterises the radial oscillation and $J_\theta$ that of oscillation in the $\theta$ direction.} where we have approximated that stars are at the position of the Sun. Changing this assumed position within our survey volume does not make any important qualitative difference. 

The distributions in Fig.~\ref{fig:actions_halo} are clearly shaped by the selection effects that apply to them. At high $|J_\phi|$ there is a lower limit for $J_r$ below which there is no orbits that pass the Sun's position. At lower $|J_\phi|$, there is a minimum $J_r$ that reaches the Sun's position for $J_\theta=0$, and this is where we find the bright maximum in the $(J_\phi,J_r)$ distribution that runs from approximately $(-1500,0)$ to $(0,600)\;\mathrm{kpc}\,\mathrm{km}\,\mathrm{s}^{-1}$, which is a selection effect. For $J_\theta>0$ there are orbits that reach the solar position at these $J_\phi$ with lower $J_r$, so the density does not fall to zero below this point. There is an upper envelope for $J_\theta$ at a given $J_\phi$, above which the density becomes very low. This is a consequence of the requirement that the orbit reaches the Sun's position and is bound to the Galaxy.

Nonetheless, there are features of the velocity distribution that stand out in these plots too. The GSE stands out as a strikingly high density of stars over a large range in $J_r$ around ${J_\phi}={0}$. ED-2 is clearly seen in both panels at $(J_r,J_\phi,J_\theta) = (600, -2450, 90)\;\mathrm{kpc}\,\mathrm{km}\,\mathrm{s}^{-1}$; the Helmi Stream is clearly visible in the $(J_\phi,J_\theta)$ plane around $(1300, 1200)\;\mathrm{kpc}\,\mathrm{km}\,\mathrm{s}^{-1}$ and expected at a $J_r$ of 80 kpc km s$^{-1}$; the Sequoia group produces an overdensity that can be seen around $(J_\phi,J_r) = (-2000,1000)\;\mathrm{kpc}\,\mathrm{km}\,\mathrm{s}^{-1}$. Otherwise, for our sample the substructure is substantially clearer in the velocity distribution than in these action distributions.

We note, in the interests of finding members of our newly discovered substructure in future study of samples beyond the Solar neighbourhood, that in our assumed potential these have actions $(J_r,J_\phi,J_\theta)$ around $(1450,200,2300)$ and $(70,1200,200)\;\mathrm{kpc}\,\mathrm{km}\,\mathrm{s}^{-1}$ for MMH-1 and MMH-2 respectively.

Finally, we can use angle-action modelling to justify an assumption underlying our approach: that it is reasonable to approximate the velocity distribution of the stellar halo as independent of position in our sample volume. To provide a realistic example, we use the Torus Code \citep{TorusCode} we sample the points from the phase-mixed orbit corresponding to the approximate actions of the Helmi and ED-2 streams. For the Helmi stream we focus only on component at $v_\theta<0$ (by symmetry the equivalent at $v_\theta>0$ will have the same spread in velocity), while for ED-2 we also limit it to $v_r<0$ to match the major component we observe. For the points on the orbital torus sampled within our survey volume, we have a dispersion in $v_\phi$ of $12\,\mathrm{km}\,\mathrm{s}^{-1}$ and in $v_\theta$ of $17\,\mathrm{km}\,\mathrm{s}^{-1}$ for the Helmi Stream, and of $(50, 40, 30)\,\mathrm{km}\,\mathrm{s}^{-1}$ in $(v_r,v_\phi,v_\theta)$ for ED-2.  This spread is comparable to that seen in for these groups in Fig.~\ref{fig:fv_halo}, which is quite small on the scale of the velocity distribution that we are studying, and clearly does not prevent us from finding substructure. We note also that this is likely to be an overestimate of the associated dispersion, because the real sample has a smaller spread in position, being preferentially near the Sun, and this smaller spread in position, for a given orbital torus, corresponds to a smaller spread in velocity.

%% file: sections/conclusions.tex
We use DR3 astrometry data without radial velocities, giving us access to a significantly larger catalogue of stars. With the penalized maximum likelihood algorithm implemented in \citetalias{paper2} we can then infer full 3D velocity distributions to investigate the Solar neighbourhood for substructure. We analyse the extended Solar neighbourhood in two separate stellar components: the Galactic disc ($d< 200$ pc) and the stellar halo ($d < 3$ kpc). The disc is also split into a north and southern hemisphere based on Galactic latitude as in \cite{monari} and we find that the overall velocity distribution is dominated by the four major moving groups; \textit{Sirius}, \textit{Coma Berenices}, \textit{Hyades}, and \textit{Pleiades}, with 35\% of stars lying in and around them. However, we find some degree of asymmetry with Galactic latitude with \textit{Coma Berenices} being most prominent in the southern hemisphere in agreement with previous results \citep{Quillen18, monari, bernet}. We also identify a new structure at $(U, V) = -10, -15\ \mathrm{km s}^{-1}$ which does not align with any known moving groups.

For the local stellar halo, we use the same approach as \cite{dr2HR} to reveal a double main sequence for stars with $v_\mathrm{T} >-200\ \textrm{km s}^{-1}$ which we split into in an \textit{`in-situ'} and \textit{`accreted'} population to the right and left in the CMD respectively. These samples are then used to infer the velocity distributions in spherical Galactocentric coordinates, $v_r, v_\phi$, and $v_\theta$. We see that we can reliably make out several of the more well-known features of the stellar halo in the \textit{`accreted'} sample: \textit{GSE}, \textit{Sequoia}, \textit{Helmi streams}, and \textit{Thamnos} are all visible in our sample. We also find three additional structures, the first of which is identified already as \textit{ED-2} in \cite{dodd}. We then have a new structure, \textit{MMH-1}, appear at large $v_\theta$, split into two locations at $(v_r, v_\theta) = (-250, 300) \ \textrm{km s}^{-1}$ and $(v_r, v_\theta) = (200, 350)\ \textrm{km s}^{-1}$. It also appears at $(v_\phi, v_\theta) = (25, 300)\ \textrm{km s}^{-1}$. By inspection of the full 3D velocity space we confirm this as one feature with velocities $(v_r, v_\phi, v_\theta) = (\pm 225, 25, 325)$ km s$^{-1}$. Lastly we also have the new feature \textit{MMH-2} at $(v_\phi, v_\theta) = (150, -100)\ \textrm{km s}^{-1}$ which we trace into $v_r$, which gives it velocities $(v_r, v_\phi, v_\theta) = (0, 150, -125)$ km s$^{-1}$. These velocity distributions gives us the appearance of the stellar halo at `face value', which provides a clear idea of what structure can be expected there and to what extent.

In addition to this, we investigated the conditional velocity distributions which provides further support for the existence of these structures and their extent. This is where we also find a match for \textit{MMH-2} in $v_r$, which was not as readily apparent in the standard distributions.

Furthermore, we transform our velocity distributions into action space distributions and identify the location of several of the previous substructures there as well. This further demonstrates the possibilities of our approach and lets us connect between velocity space and orbital space.

A promising benefit of these velocity distributions is the ability to produce stellar candidate lists of velocity features. While the method obviously does not allow us to determine the $v_r$ of individual stars, it does allow for determining a probability of having the necessary $v_r$ to belong to a certain velocity feature. Consider again our 3D grid of velocities from $\bm{v}_min$ to $\bm{v}_max$ with some grid spacing. In this box, a star with unknown $v_r$ forms a line $\bm{p} + v_r\bm{\hat{r}}$. If we make the reasonable assumption that the true $v_r$ lies within our box, then we can determine the integral of the probability distribution where the line crosses the velocity feature. This is then normalised against the probability distribution along the entire line:
\begin{equation}
    P_k({\bm{l^*}}) = \frac{\mathlarger{\sum}\limits_{\bm{l^*}} e^{\phi_{\bm{l^*}}} K(k|{\bm{l^*}}) }
    {\mathlarger{\sum}\limits_{\bm{l}} e^{\phi_{\bm{l}}} K(k|\bm{l})},
\end{equation}
where $\bm{l^*}$ are the cells assigned to a specific velocity feature (where cells $\bm{l}$ are defined in Section \ref{sec:theory}). Once this is determined, a candidate list is created by requiring that $P_k(\bm{l^*})$ be greater than some threshold probability.

This work demonstrates what can be achieved without needing to rely on the full 6D phase-space information. During the era of \textit{Gaia}, there will be more sources with astrometry alone than with added radial velocities due to the inherent differences in the methods by which the measurements are obtained. Currently around 2\% of the data has radial velocities and with spectroscopic follow-up this is likely to increase, but not match the amount of pure astrometric sources. In the next era with a successor mission in the infrared (Hobbs 2022, submitted)\footnote{Proceedings of the XXXI IAU General Assembly, to be published in Cambridge University Press} the amount of radial velocities could increase significantly. This will mean that enough sources will be available that more discoveries can be made directly with 6D data and in our case it could allow us to determine the $\alpha$-parameter more finely, but the proper motions will remain more numerous and so methods such as these will be pivotal. Additionally, as the number of 5D sources increase, the velocity resolution that can be achieved in the estimation of the velocity distribution will also increase.